\documentclass[pra,showpacs,floatfix,amsmath,amsfonts,twocolumn]{revtex4}

\usepackage{graphics}
\usepackage{epsfig}
\usepackage{color}
\usepackage{eurosym}
\usepackage[normalem]{ulem}
\usepackage[makeroom]{cancel}

\newcommand {\tpsi}{\tilde{\psi}}
\newcommand {\tk}{{k}}

\newcommand{\p}{{\cal P}}
\newcommand{\PT}{{\cal PT}}
\newcommand{\T}{{\cal T}}

\newcommand{\bphi}{\boldsymbol{\phi}}
\newcommand{\sech}{\mathrm{sech}}

 \newcommand{\lstar}{\lambda_\star}
\newcommand{\kstar}{k_\star}

\begin{document}

\title{Phase transition through the splitting of self-dual spectral singularity in optical potentials}

\author{Vladimir V. Konotop$^{1,*}$ and Dmitry A. Zezyulin$^{2,3}$}

\affiliation{$^1$Centro de F\'isica Te\'orica e Computacional and Departamento de F\'isica, Faculdade de Ci\^encias,
	Universidade de Lisboa, Campo Grande 2, Edif\'icio C8, Lisboa 1749-016, Portugal\\
$^2$ITMO University, St. Petersburg 197101, Russia\\
$^3$Institute of Mathematics with Computer Center, Ufa Scientific Center,
	Russian Academy of Sciences, Chernyshevskii str., 112, Ufa 450008, Russia}

\date{\today}

\email{vvkonotop@fc.ul.pt }

\begin{abstract}
We consider optical media which feature antilinear symmetries. We show that: (i) spectral singularities of such media (if any) are always self-dual, i.e., correspond to  CPA-lasers; (ii) under the  change of a  system's  parameter the self-dual spectral singularity may split into a pair of isolated complex conjugate eigenvalues,  which corresponds to an unconventional and overlooked in the most of previous studies scenario of the phase transition (known as $\PT$-symmetry breaking in systems obeying parity-time symmetry); (iii) if the antilinear symmetry is local, i.e., does not involve any spatial reflection, then no spectral singularity is possible. Our findings are illustrated with several examples including a  $\PT$-symmetric bilayer
and other complex potentials discussed in recent literature.
\end{abstract}


\maketitle

One of the most fascinating features of a typical parity-time ($\PT$-) symmetric system is its ability to undergo the phase transition,  which corresponds to the qualitative change of optical properties  of the medium.  The phase transition is associated with   transformation of the spectrum of a system from all-real  (the phase of unbroken $\PT$ symmetry) to  partially complex (phase of   spontaneously broken $\PT$ symmetry)~\cite{Bender,BenBoet}. This transition was observed in pioneering experiments~\cite{Christod,Kip}, as well as in numerous subsequent studies (see  \cite{KYZ} for review). In the meantime, $\PT$-symmetric optical media constitute just a subclass of a more general family  of pseudo-Hermitian  systems which are described by Hamiltonians  commuting with antilinear operators \cite{Mostafazadeh2002,Solombrino}.  Complex eigenvalues of  a pseudo-Hermitian Hamiltonian (if any)  always emerge as complex conjugate pairs. This restriction allows for generalization of the concept of unbroken and broken phases from the context of $\PT$ symmetry to much broader realm of pseudo-Hermitian optical media. Phase transitions from all-real to complex spectra in several classes of complex non-$\PT$-symmetric potentials associated with various antilinear symmetries have been studied recently \cite{NY16,Yang}.

A common  mechanism of transition from entirely real to complex spectrum is the occurrence of an exceptional point (EP)~\cite{BenBoet,Bender}. The notion of EP was introduced in mathematics several decades ago~\cite{Kato}; an EP  corresponds to a situation when continuous change of some control parameter $\lambda$, characterizing a system, leads  to the coalescence of two isolated eigenvalues (i.e., propagation constants) at some critical value $\lambda_{EP}$.   Moreover, the coalescence of eigenvalues is accompanied by the merge of the associated eigenvectors, which means that the system Hamiltonian becomes nondiaginazable at the EP. As the control parameter is driven beyond the critical value $\lambda_{EP}$, the double eigenvalue splits as a complex conjugate pair, and the system  enters the phase of spontaneously broken symmetry (for a more detailed discussion see e.g. \cite{Rotter}). 

The collision of two isolated eigenvalues is not the only possible mechanism of the phase transition.
 For instance, an  exotic scenario of $\PT$-symmetry breaking  was reported on in a spinor $\PT$-symmetric system \cite{KKZ}. Very recently, an unconventional mechanism of the phase transition  was encountered in the numerical study in~\cite{Yang}. It was observed that a pair of complex conjugate eigenvalues spontaneously bifurcates {\em from an interior point of the continuous spectrum} of a specific optical potential; moreover, this complex conjugate pair bifurcates out from continuous (non-square-integrable) eigenmodes (rather than from the bound states embedded in continuum~\cite{BIC}). This phase transition however has  not received an explanation yet and was left by the author as a ``mathematical mystery''. The goal of the present Letter, is to give an explanation for  the observed phenomenon and show that this is a common phenomenon for optical systems which at certain parameters obey self-dual singularities.

Consider the wave diffraction in the paraxial approximation (in dimensionless units)
\begin{eqnarray}
\label{paraxial}
i\Psi_z+\Psi_{xx}+V(x)\Psi=0,
\end{eqnarray}
with a complex optical potential $V(x)$ which is spatially localized: $V(x)\to 0$ at $|x|\to\infty$. 
Stationary modes, $\Psi(z,x)=e^{ibz}\psi(x)$ solve  the dimensionless Helmholtz equation
\begin{eqnarray}
\label{Helmholtz}
H\psi=b\psi, \qquad H =\partial_{xx}+V(x), 
\end{eqnarray} 
where $b$ is the propagation constant. If  at $|x|\to \infty$, the potential $V(x)$ decays fast enough, then the spectrum of $b$ consists of all-real continuum which fills the semiaxis $b<0$ and bound states with localized eigenfunctions. Bound states (if any)  can have real or complex propagation constants. If all propagation constants of bound states are real, then the phase is unbroken; presence of complex propagation constants in the spectrum corresponds to the broken phase.


Now we recall that the continuous spectrum may contain special points  which are known as {\em spectral singularities}. This concept, known in mathematical literature for long time~\cite{Neimark},  has recently acquired the increasing interest   due to 
unusual physics associated to spectral singularities~\cite{Mostafazadeh2009,Laser1,Laser2}, which includes the
idea and experimental implementation of the coherent perfect absorber (CPA)~\cite{Stone,Wan} (which can be viewed as a time-reversed spectral singularity), as well as the idea of a CPA-laser~\cite{Longhi,Mostafazadeh2012}. 

Since a spectral singularity belongs to the continuous spectrum of a potential $V(x)$, it is convenient to parametrize propagation constants as $b = -k^2$, where $k$ is a new parameter. Real and nonnegative values of $k$ correspond to the continuous spectrum of propagation constants on the  semi-axis $b\leq 0$.   
Now, consider two pairs of the Jost solutions of (\ref{Helmholtz})  defined uniquely by their asymptotics to the left (``L'') and right (``R'') from the potential:
\begin{eqnarray}
\label{Jost}
\begin{array}{ccc}
\phi_1^{L}(x)\to e^{ikx} & \phi_2^{L}(x)\to e^{-ikx} & \mbox{at $x\to-\infty$,} 
\\
\phi_1^{R}(x)\to e^{ikx} & \phi_2^{R}(x)\to e^{-ikx} & \mbox{at $x\to+\infty$,} 
\end{array}
\end{eqnarray}
and introduce the transfer matrix $M$ through the relation
\begin{eqnarray}
\label{M}
\bphi^R=M\bphi^L, \quad\mbox{where}\quad \bphi^{L,R} = \left(\!\!\begin{array}{c}
\phi_1^{L,R}\\ \phi_2^{L,R}\end{array}\!\!\right).
\end{eqnarray}
 
Let the potential    depend on a parameter, say $\lambda$, i.e.,  $V(x)\equiv V(x;\lambda)$. Correspondingly, the transfer matrix $M$ depends of $k$ and $\lambda$. If for some $\lambda$, the entry $M_{22}$ of the transfer matrix vanishes   at some value of $k_2>0$ [i.e., $M(k_2; \lambda)=0$], then  $k_2$ is called the {\em spectral singularity} of  potential $V(x; \lambda)$.  Similarly, if $M_{11}(k_1; \lambda)=0$ at $k_1>0$, the value $k_1$ is called  {\em time-reversed spectral singularity} of $V(x; \lambda)$.  Next, if for some value $\lambda=\lambda_\star$ the spectral singularity   and the time-reversed spectral singularity coincide, i.e., $k_1=k_2=k_\star$ and $M_{11}(k_\star; \lambda_\star) = M_{22}(k_\star; \lambda_\star)=0$,  then $k_\star$ is called   {\em self-dual spectral singularity} for potential $V(x; \lambda_\star)$ \cite{Mostafazadeh2012}. 
We also note that a zero of $M_{22}(\lambda,\tk_2)$ with Im\,$\tk_2>0$, as well as a zero $M_{11}(\lambda,\tk_1)$ with Im\,$\tk_1<0$, correspond to {\em discrete} eigenvalues, i.e., to bound states with complex propagation constants. Now we can formulate the main results of this Letter, describing a novel mechanism of the phase transition, in general, and, in particular, explaining the numerical findings in \cite{Yang}.

Let a potential $V(x;\lambda)$ vanish  at the infinities $x\to \pm\infty$ and  have entirely real spectrum. 
Let at some $\lambda$ the zeros of $M_{11}$ and $M_{22}$ be   located in the upper and lower half-planes of complex $k$, i.e., Im\,$\tk_1>0$ and Im\,$\tk_2<0$. As the parameter $\lambda$ is being changed, the dependencies $\tk_1(\lambda)$ and $\tk_2(\lambda)$ evolve in the complex plane $k$ and eventually cross the real axis. This crossing occurs at $\lambda=\lstar$ in a self-dual spectral singularity $\kstar=\tk_1(\lambda_\star)=\tk_2(\lambda_\star)$. Further continuous change of $\lambda$ results in passing the roots $\tk_j$ to the opposite half-planes of the complex $k$: i.e., it results in  Im\,$\tk_1<0$ and Im\,$\tk_2>0$, which  corresponds to the emergence of two isolated complex eigenvalues with spatially localized eigenfunctions. Therefore, the system undergoes the transition  from the entirely real spectrum to the complex spectrum  trough   {\em splitting of a self-dual spectral singularity} as the control parameter is being changed.

In the above discussion we  have not imposed any particular restriction on the potential $V(x;\lambda)$. However it is natural to expect that the described mechanism is more likely to be encountered in $\PT$-symmetric systems, because for $\PT$-symmetric potentials spectral singularities and time-reversed spectral singularities occur simultaneously  (i.e., at the same $k=k_\star$)~\cite{Longhi}. Remarkably, in spite of intensive study of $\PT$ symmetry in recent decade, splitting of self-dual singularities was mentioned in only a few previous works on $\PT$-symmetric systems~\cite{BQ2010, HHK} and has not received any attention in  the context of phase transitions.  

\begin{figure}
	\includegraphics[width=1.0\columnwidth]{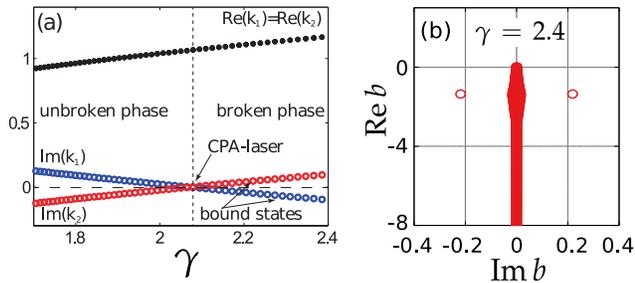}
	\caption{(a) Real and imaginary  parts of the roots $k_1$ and $k_2$ of the transfer matrix elements $M_{11}$ and $M_{22}$ for   the potential (\ref{eq:bilayer}) as functions of the gain-and-loss coefficient $\gamma$. Real parts of $k_1$ and $k_2$ always coincide. Self-dual spectral singularity  (i.e., CPA-laser) occurs at $\gamma_{\star}\approx 2.072$ and $k_\star \approx  1.065$~\cite{Mostafazadeh2009}. 		The vertical dashed line  demarcates  unbroken phase, $\gamma<\gamma_{\star}$, and broken phase  $\gamma>\gamma_{\star}$.
		   In the broken phase,  the zeros $k_{1,2}$ correspond to the bound states with complex propagation constants.  (b) Numerically obtained spectrum of propagation constants for $\gamma=2.4>\gamma_{\star}$. A pair of isolated bound states with complex-conjugated propagation constants is visible, which means   broken $\PT$ symmetry.}
	\label{fig:one}
\end{figure}

Let us first illustrate the described transition using a simple model of a 
 $\PT$-symmetric bilayer consisting of two slabs of the unit width:
\begin{eqnarray}
\label{eq:bilayer}
V(x; \gamma)=\left\{ \begin{array}{cc}
+i\gamma & x\in (-1,0)
\\
-i\gamma & x\in (0,1)\\
0 & \mbox{otherwise.}
\end{array}\right.  
\end{eqnarray}
Here $\gamma >0$ is the  control parameter which describes  gain in the  active medium $x\in (-1,0)$ and losses in the absorbing  layer $x\in (0,1)$. Spectral singularities of this potentail are described in~\cite{Mostafazadeh2009}.
It is straightforward to compute the transfer matrix $M(k; \gamma)$ of (\ref{eq:bilayer}) explicitly, but the corresponding expression is too bulky to be presented here. Instead, in Fig.~\ref{fig:one}   we illustrate the ``dynamics'' of zeros of $M_{11}$ and $M_{22}$ as  the gain-and-loss coefficient $\gamma$ changes. 
As $\gamma$ increases, the complex roots $k_1$ and $k_2$ meet at the real axis at the self-dual spectral singularity  which corresponds to   $\gamma_{\star}\approx 2.072$  and $\kstar\approx 1.065$ \cite{Mostafazadeh2009}. The    increase of $\gamma$ beyond $\gamma_{\star}$ leads to the splitting of the self-dual spectral singularity into a pair of isolated bound states with complex-conjugated propagation constants. Thus the transition of the control parameter through the self-dual spectral singularity corresponds to the $\PT$-symmetry breaking. In order to verify this conclusion, we have  computed numerically the entire spectrum of potential $V(x; \gamma)$ for several values of $\gamma$. For $\gamma<\gamma_{\star}$, the spectrum is entirely real and   continuous (not shown in Fig.~\ref{fig:one}). However, for $\gamma>\gamma_{\star}$  [Fig.~\ref{fig:one}(b)], a pair of complex-conjugated propagation constant emerges from the interior of the continuum, i.e., the phase transition indeed takes place. 

Thus the described splitting scenario is typical for $\PT$-symmetric systems, where self-dual spectral singularities  are ubiquitous. It is, however, not necessary for a system to be $\PT$ symmetric in order to obey self-dual spectral singularities~\cite{Mostafazadeh2012}. Moreover,  $\PT$-symmetric systems constitute just a subclass of a more general  family of pseudo-Hermitian systems described by the Hamiltonians (i.e. by $H$ in (\ref{Helmholtz})) commuting with antilinear operators  (in the $\PT$-symmetric case this antilinear operator is the $\PT$ operator, i.e., the combination of parity $\p$ and time $\T$ reversals). In what follows, we use this idea in order to demonstrate that the phase transition through the splitting of a spectral singularity can be encountered in a broad class of potentials which commute with an antilinear symmetry involving the spatial reflection (including $\PT$ symmetry as a particular case).  It is exactly this phase transition  that was observed in \cite{Yang}.  Additionally, we demonstrate that if the antilinear symmetry is ``local'', i.e., if it does not involve spatial inversion, then the potential cannot support a spectral singularity.

We focus on optical media described by Hamiltonians $H$ in (\ref{Helmholtz}) for which one can find an invertible intertwining  operator $\eta$ satisfying the condition  
\begin{eqnarray}
H^\dagger=\eta H\eta^{-1}.
\end{eqnarray}
These are pseudo-Hermitian~\cite{Mostafazadeh2002} operators if $\eta$ is Hermitian, and weakly  pseudo-Hermitian ones for all other invertible $\eta$~\cite{Solombrino} (to simplify the terminology,  below we use the term ``pseudo-Hermitian'' for any $\eta$). Pseudo-Hermitian operators include $\PT$-symmetric ones as a subclass with operator $\eta$ being the parity operator~$\p$~\cite{Mostafazadeh2002}. For stationary solutions governed by Eq.~(\ref{Helmholtz}), $\p$ is the spatial reflection, $\p\psi(x) = \psi(-x)$, and $\T$ is the complex conjugation, $\T\psi = \psi^*$. 
 
For operator $H$ in (\ref{Helmholtz}), $H^\dag = H^*$. Hence, if $\psi$ is an eigenfunction of (\ref{Helmholtz}) corresponding to the real    $b$, then   $\tpsi= (\eta  \psi)^* = \T \eta \psi$ is also an eigenfunction with the same propagation constant $b$. While the intertwining operator $\eta$ may depend on the spatial coordinate (see e.g. the example (\ref{Yang}) below), in this Letter we consider a special (but still very broad) class of operators $\eta$  whose action  for $x\to \infty$ and $x\to -\infty$ can be approximated by  some operator $\eta_\infty$ which is the same for $+\infty$ and $-\infty$; in other words, 
we define $\eta_\infty=\lim_{x\to+\infty}\eta=\lim_{x\to-\infty}\eta$. 

Consider the potential~\cite{Yang}  
\begin{eqnarray}
\label{Yang}
V(x)=h'(x)-h^2(x), \qquad \lim_{x\to\pm \infty}h(x)=0,
\end{eqnarray}
where the prime denotes the derivative with respect to $x$ and $h(x)$ is an arbitrary $\PT$-symmetric function. Generally speaking, potentials (\ref{Yang})   are not $\PT$ symmetric; however  they correspond to  pseudo-Hermitian Hamiltoninans with the intertwining operator~\cite{Yang}: $\eta=\p [\partial_x +h(x)]$  and  $\eta_\infty=\p\partial_x$.

Returning to the Jost solutions $\bphi^L$ and $\bphi^R$ in  (\ref{Jost}), we note that in view of the presence of the antilinear symmetry, functions $\T\eta \bphi_{1,2}^L$ and $\T\eta \bphi_{1,2}^R$ also solve Eq.~(\ref{Helmholtz}) with the same $k>0$.  Since for large $x$ operator $\eta$ can be approximated by $\eta_\infty$ and since the Jost solutions  are uniquely defined by their asymptotics, we find that left and right solutions are connected through the following transformations:
\begin{equation}
\label{aux}
\T\eta \bphi^L = -ik\sigma_3\bphi^R, \quad \T\eta \bphi^R = -ik\sigma_3\bphi^L \quad \mbox{for all $k>0$}
\end{equation}
(hereafter $\sigma_j$ is the standard notation for Pauli matrices).
Applying $\T\eta$ operator to  both sides of (\ref{M}) and using (\ref{aux}), we readily obtain the  following identity
\begin{eqnarray}
\label{rel_M}
M^{-1}=\sigma_3M^*\sigma_3 \qquad \mbox{for all $k>0$}.
\end{eqnarray}
This result, together with the requirement $\det M=1$,  yields the following relations between the entries of the transfer matrix:
\begin{eqnarray}
\label{entries_M}
M_{11}=M_{22}^*, \qquad  { M_{12}=M_{12}^*, \qquad   M_{21}=M_{21}^*}  
\end{eqnarray}
for all $k>0$.  These immediately imply that if  an  optical potential  of the form (\ref{Yang})  has a spectral singularity, then the latter is necessarily self-dual, i.e., $M_{11}=0$ if and only if $M_{22}=0$.

Note that the obtained relations (\ref{rel_M})--(\ref{entries_M}) between the entries  of the  transfer matrix are similar but not identical to the  analogous relations for   $\PT$-symmetric potentials ($\eta=\p$), where  the transfer matrix obeys the following identity  for $k>0$~\cite{Longhi}:  $M^{-1} = M^*$,
and respectively
\begin{eqnarray}
M_{11} = M_{22}^*, \qquad  M_{12}^* = -M_{12}, \qquad M_{21}^* = -M_{21}.
\end{eqnarray}

 Now we demonstrate that the scenario of  phase transition through the splitting of a self-dual spectral singularity takes place in a non-$\PT$-symmetric potential of the form (\ref{Yang}). More specifically, we consider  the potential $V(x)$ generated by the following function:
\begin{equation}
\label{sech}
h(x) = d_1 \sech \, x + id_2 \tanh x \, \sech x, 
\end{equation}
where $d_1$ is fixed to $1$ and $d_2$ is a varying control parameter. In \cite{Yang} it was observed that the spectrum of the corresponding potential $V(x)$ is all-real if $d_2< d_{2,\star}  \approx 1.385$, but a pair of complex eigenvalues bifurcates out from the interior of the continuous spectrum as $d_2$ exceeds the critical value $d_{2,\star}$.    

In order to explain the observed in \cite{Yang} behavior, we computed numerically the Jost solutions and the transfer matrix. In Fig.~\ref{fig:one}(a), we show $|M_{11}| = |M_{22}|$  for three different values of $d_2$. Note that the curves are plotted as functions of $k^2=-b$.  It is readily observed that at $d_2=d_{2,\star}$  (which corresponds to the blue curve)  amplitudes of $M_{11}$ and $M_{22}$ vanish simultaneously     at $b_\star=-k_\star^2 \approx -0.806$, which corresponds to the interior point of the continuous spectrum where the bifurcation of complex eigenvalues was detected in \cite{Yang}.  Thus the emergence of complex eigenvalues is explained by the transition through the self-dual spectral singularity. Additionally, in Fig.~\ref{fig:one}(b) we plot  the amplitude of the transmission coefficient $T$ as well as coefficient of left ($R^L$) and right ($R^R$) reflections for $d_{2,\star}$. One readily observe that at the propagation constant $b_\star$ corresponding to the self-dual spectral singularity all the three plotted curves diverge simultaneously. 
As an interesting byproduct, we observe that at $k^2\approx 0.522$ the coefficient $R^L$ vanishes, while $R^R$ has     nonzero value; hence the unidirectional reflectionlessness   occurs at the corresponding wavenumber.

\begin{figure}
	\includegraphics[width=\columnwidth]{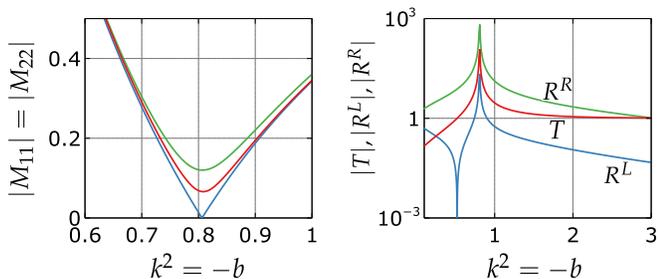}
	\caption{(a) Amplitudes of $M_{1,1}$ and $M_{2,2}$ plotted as dependencies of $k^2=-b$ for the potential (\ref{Yang}), (\ref{sech}) for $d_1=1$ and   three different values of $d_2$: 1.35 (red line),  $d_{2,\star} = 1.385$ (blue line) and 1.45 (green line). (b) Logarithmic plot of transmission and reflection coefficients for $d_{2,\star}$.}
	\label{fig:two}
\end{figure}

The notion of the antilinear symmetry 
also proves to be useful in order to describe   classes of complex potentials where spectral singularities are   forbidden. Indeed, let us consider the class of  localized potentials \cite{Andrianov,Wadati} 
 \begin{equation}
 \label{Wadati}
 V(x) = g^2(x) + i g'(x),  \quad \lim_{x\to\pm \infty} g, g'(x) = 0,
 \end{equation}
 where $g(x)$ is a real-valued function (no symmetry for $g(x)$ is assumed).
Now the intertwining operator has the form \cite{NY16} $\eta = \partial_x + ig(x)$. Thus if $g$ and $g'$ decay  rapidly enough, $\eta_\infty=\partial_x$. Then, applying the same strategy as above, we find
 \begin{equation}
 \label{MW}
 M^* = \sigma_1 M \sigma_1,  \quad \mbox{for all $k>0$}.
 \end{equation}
 i.e., 
 \begin{eqnarray}
 \label{entries_MW}
 M_{11}(k)=M_{22}^*(k),  \quad M_{12}(k)=M_{21}^*(k)\quad \mbox{for all $k>0$}.
 \end{eqnarray}
 Thus  for the entries of transfer matrix of  potential (\ref{Wadati})  we obtained the same relations as for any Hermitian potential, i.e., the scattering behavior in these complex potentials has some distinctive traits of Hermitian behavior. In particular,   {\em no spectral singularities is possible in potentials} (\ref{Wadati}): indeed, if $M_{11}=M_{22}=0$, then relations (\ref{entries_MW}) trivially imply that $\det M \leq 0$, which contradicts to the basic requirement $\det M=1$. The  same conclusion holds for another class of localized complex potentials in the form \cite{NY16,Andrianov}
 \begin{equation}
 \label{Andrianov}
 V(x) = \frac{1}{4} g^2 + \frac{g'^2 - 2 gg'' +c}{4g^2} + ig',
 \end{equation}
 where $c$ is a real constant and $g(x)$ is a real-valued localized function. For this family of potentials, $\eta$ is the second-order differential operator $\eta=\partial_{xx} + \alpha_1(x)\partial_x + \alpha_0(x)$ whose coefficients $\alpha_{0,1}(x)$ are localized functions which  can be expressed through $g$ \cite{NY16};  then $\eta_\infty = \partial_x^2$, and relations (\ref{MW}) and (\ref{entries_MW}) remain valid.
 
 Several examples considered above help us to develop simple intuition about the relation between an antilinear symmetry and the presence of spectral singularities: if the antilinear symmetry is ``nonlocal'', i.e., involves spatial reflection $\p$, then spectral singularities can occur and they are necessarily self-dual.  This is the case of $\PT$-symmetric potentials and of (generically non-$\PT$-symmetric) potentials (\ref{Yang}). However, if the antilinear symmetry is ``local'', i.e., does not involve any spatial reflection, the spectral singularities are forbidden. This is what happens for 
 potentials (\ref{Wadati}) and   (\ref{Andrianov}). Another distinctive feature of potentials  (\ref{Wadati}) and  (\ref{Andrianov}) is the interdiction of unidirectional reflectionless or invisibility. Indeed, $M_{12}=0$ if and only if $M_{21}=0$, i.e., invisibility (if any) is always bidirectional.

  To conclude, stimulated by recent numerical observation reported on in~\cite{Yang}, we uncover a mechanism of phase transition  between a  purely real spectrum of an optical system and a spectrum possessing complex eigenvalues. This mechanism consists in splitting a self-dual singularity into a pair of isolated complex-conjugate eigenvalues.
  The described phase transition is fairly general and is typical to occur in $\PT$-symmetric and pseudo-Hermitian potentials where spectral singularities are always self-dual, i.e., correspond  to CPA-lasers.
  On the other hand, if the antilinear symmetry is local, i.e., does not involve any spatial reflection, then no spectral singularity is possible, i.e., the described scenario cannot be implemented. 
  
Finally we recall, that until now one common feature of exceptional points and spectral singularities has been known: both these phenomena are associated with the situation when the eigenfunctions of the Hamiltonian loses their completeness (i.e., do not form a complete basis). In this Letter, we have shown that there also exits   a similarity in the phase transitions caused by the presence of exceptional points and spectral singularities: in both cases one observes splitting of the spectrum and bifurcation of bound states towards the complex plain; in case of exceptional points this usually occurs in the discrete spectrum, while the bound states born by   spectral singularities emerge from   interior point of the  continuum.

  \bigskip

  The research of D.A.Z.  is supported by Russian Science Foundation (grant no. 17-11-01004).

\end{document}